\documentclass[traditabstract]{aa} 
\usepackage{multirow} 
\usepackage{natbib}
\usepackage{amssymb,amsmath}
\usepackage{graphicx, subfigure}
\usepackage{siunitx}
\usepackage{txfonts}
\usepackage{url}
\usepackage{threeparttable}
\usepackage[table]{xcolor}
\usepackage{makecell}
\usepackage{epstopdf}
\begin{document}
   \title{Study of the inner dust envelope and stellar photosphere of the AGB star R Doradus using SPHERE/ZIMPOL}

   \author{T. Khouri\inst{1,2}\thanks{{\it Send offprint requests to T. Khouri}\newline \email{theokhouri@gmail.com}}, M. Maercker\inst{1}, L. B. F. M. Waters \inst{2,3},
   W. H. T. Vlemmings\inst{1}, P. Kervella \inst{4,5}, A. de Koter \inst{2}, C. Ginski \inst{6}, E. De Beck \inst{1}, L. Decin \inst{7}, M. Min \inst{2,3} C. Dominik \inst{2},
   E. O'Gorman \inst{1}, H.-M. Schmid \inst{8}, R. Lombaert \inst{1}, E. Lagadec \inst{9}
}

\institute{Department of Earth and Space Sciences, Chalmers University of Technology, Onsala Space Observatory, 439 92 Onsala, Sweden 
\and Astronomical Institute ``Anton Pannekoek", University of Amsterdam, PO Box 94249, 1090 GE Amsterdam, The Netherlands 
            \and
            SRON Netherlands Institute for Space Research, Sorbonnelaan 2, 3584 CA Utrecht, The Netherlands 
            \and
Unidad Mixta Internacional Franco-Chilena de Astronom\'{i}a (CNRS UMI 3386), Departamento de Astronom\'{i}a, Universidad de Chile, Camino El Observatorio 1515, Las Condes, Santiago, Chile 
\and
LESIA (UMR 8109), Observatoire de Paris, PSL, CNRS, UPMC, Univ. Paris-Diderot, 5 place Jules Janssen, 92195 Meudon, France 
\and
	Sterrewacht Leiden, P.O. Box 9513, Niels Bohrweg 2, 2300RA Leiden, The Netherlands 
	\and
       Instituut voor Sterrenkunde, KU Leuven, Celestijnenlaan 200D B-2401, 3001 Leuven, Belgium 
       \and
       Institute for Astronomy, ETH Zurich, 8093 Zurich, Switzerland 
       \and
       Laboratoire Lagrange, Universit\'e C\^ote d'Azur, Observatoire de la C\^ote d'Azur, CNRS, Blvd de l'Observatoire, CS 34229, 06304 Nice cedex 4, France 
}

  \abstract
  {On the asymptotic giant branch (AGB) low- and intermediate-mass stars eject a large fraction of their envelope,
  but the mechanism driving these outflows is still poorly understood.
  For oxygen-rich AGB stars, the wind is thought to be driven by radiation pressure
  caused by scattering of radiation off dust grains.
  We use high-angular-resolution images obtained with SPHERE/ZIMPOL to study the photosphere, the warm molecular layer, and the inner wind
  of the close-by oxygen-rich AGB star R~Doradus and its inner envelope.
  We present observations in filters V, cntH$\alpha$, and cnt820
  and investigate the surface brightness distribution of the star and of the polarised light produced in the inner envelope.
  Thanks to second-epoch observations in cntH$\alpha$, we are able to see variability on the stellar photosphere.
  We study the polarised-light data using a continuum-radiative-transfer code that accounts for direction-dependent scattering of photons off dust grains.
  We find that in the first epoch the surface brightness of R~Dor is asymmetric in V and cntH$\alpha$, the filters where
  molecular opacity is stronger, while in cnt820 the surface brightness is closer to being axisymmetric.
  The second-epoch observations in cntH$\alpha$ show that the morphology of R~Dor
  has changed completely in a timespan of 48 days to a more axisymmetric and compact configuration. This variable morphology is probably linked to
  changes in the opacity provided by TiO molecules in the extended atmosphere.
  The observations show polarised light coming from a region around the central star. The inner radius of the region from where polarised light is seen varies
  only by a small amount with azimuth. The value of the polarised intensity, however, varies by between a factor of 2.3 and 3.7 with azimuth for the different images.
  We fit the radial profile of the polarised intensity using a spherically symmetric model
  and a parametric description of the dust density profile, $\rho(r) = \rho_\circ r^{-n}$. On average, we find exponents
  of $- 4.5 \pm 0.5$ that correspond to a much steeper density profile than that of a wind expanding at constant velocity.
 The dust densities we derive imply an upper limit for the dust-to-gas ratio of $\sim 2\times10^{-4}$ at 5.0\,$R_\star$.
Considering all the uncertainties in observations and models,
this value is consistent with the minimum values required by wind-driving models for the onset of a wind, of $\sim 3.3\times10^{-4}$. However, if the steep
density profile we find extends to larger distances from the star, the dust-to-gas ratio will quickly become too small for the wind of R~Dor
to be driven by the grains that produce the scattered light.}
               
\titlerunning{A study of R Dor using ZIMPOL}
\authorrunning{T. Khouri et al.}

\maketitle
%

\section{Introduction}

The asymptotic giant branch (AGB) is one of the final stages of the evolution of low- and intermediate-mass
stars, when a slow and dense outflow develops \citep{Habing2003}.
The wind is thought to be driven by radiation pressure on dust grains that can only form because pulsations enhance the density-scale-height of
the stellar atmospheres.
For oxygen-rich AGB stars (where the carbon-to-oxygen ratio is lower than one), it has been proposed that large, translucent dust grains
provide the required opacity and drive the wind through scattering of photons \citep{Hofner2008}.
However, many intricacies of the formation and processing of the oxygen-rich dust grains remain poorly constrained from observations.
The translucent nature of these grains implies that they are not expected to produce significant infrared emission and, hence, most likely cannot be identified
from infrared spectra. The best way to study such grains is through the scattered stellar light they are expected to produce.

To advance our understanding of the AGB mass loss, we use high-angular-resolution observations acquired with SPHERE/ZIMPOL \citep{Beuzit2008}
on the Very Large Telescope (VLT) to investigate light polarised through scattering off dust grains
in the inner wind of the AGB star R\,Doradus.
This oxygen-rich star with spectral type M8 has a large angular diameter in the sky ($\approx 57$\,mas) and a relatively low mass-loss rate, between
$0.9~{\rm and}~2.0 \times 10^{-7}$~M$_\odot$~yr$^{-1}$ \citep{Olofsson2002,Maercker2008,KhouriPhD}.
Its pulsation properties switch between one mode with a period of 332 days and ${\Delta V = 1.5\,{\rm mag}}$,
and another with a period of 175 days and ${\Delta V < 1\,{\rm mag}}$ \citep{Bedding1998}.
Polarised light from a region very close to the star ($\approx 1.5\,R_\star$) has been recently detected
using NACO \citep{Norris2012}. The authors found that a model with grain radii of 0.3\,$\mu$m
and a inner radius of the dust envelope of 43.3 mas gives the best fit to the data.

\section{Observations}

 \begin{figure*}[t]
   \centering
      \includegraphics[width= 18cm]{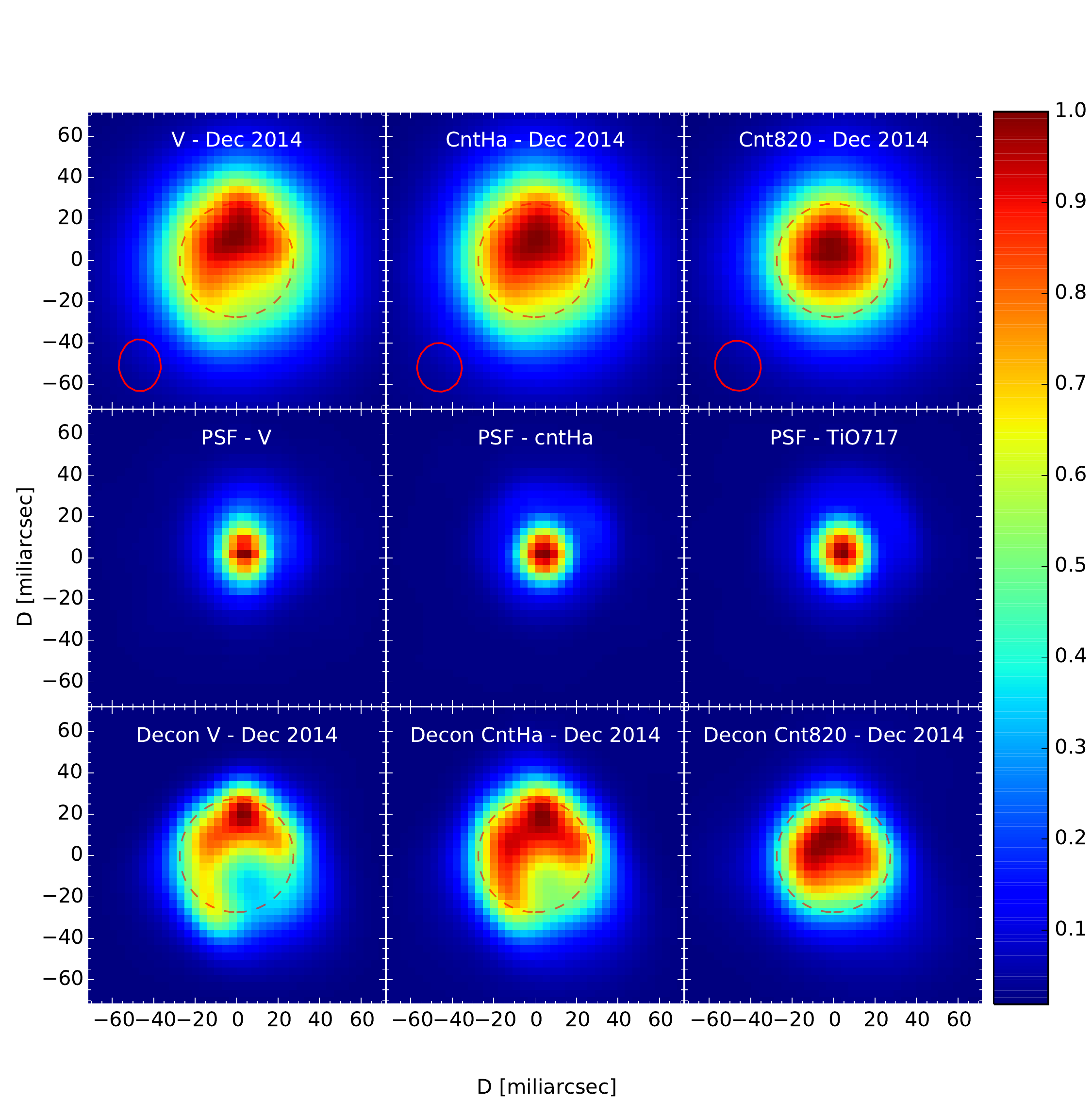}
      \caption{{\it Top panels}: total intensity normalised using the peak flux as observed using ZIMPOL on 10- and 11-Dec-2014 in the three filters,
      V, cntH$\alpha$, and cnt820. {\it Middle panels}: the images of the PSF-reference star $\psi^2$~Ori normalised to the peak flux.
      {\it Bottom panels}:
      the corresponding deconvolved images, again normalised using the peak flux. The dashed red circles show the size of the stellar disc
      derived by \cite{Norris2012} from observations in the near-infrared (same as shown in Figs. \ref{fig:cntHa}, \ref{fig:filters_contours},
      \ref{fig:polangle}, and \ref{fig:polDeg}). The full red circles show the FWHM of the PSF used as reference.}
         \label{fig:filters}
   \end{figure*}

\begin{table*}
\footnotesize
\setlength{\tabcolsep}{4pt}
\caption{Observation log.}              
\label{tab:obs}      
\centering                                      
\begin{tabular}{r c c c c c c c c c c c c}
\hline\hline                        
Filter & Average JD & Exp time & DIT[s] & $\lambda_\circ$ & $\Delta \lambda$ & AM & $\theta$ & FWHM$_\star$ & Peak PD & MPF & IPF & Rem. \\    
& [2457000+] & [min] & x NDIT & [nm] & [nm] & & [$''$] & [$mas$] & [\%] & [\%] &  [\%] & \\
\hline                                   
    V - 11 Dec 14 & \phantom{1}2.563 & 42.7 &\phantom{1.}4 x 10 & 554.0 & 80.6 &1.40 & 1.0\phantom{0} & 70.6 & 4.8 & \phantom{0}$1.85 \pm 0.15$ & $1.25 \pm 0.1$\phantom{0} & - \\ 
    cntH$\alpha$ - 10 Dec 14 & \phantom{1}1.767 & 48.0 & \phantom{.}10 x 4\phantom{6} & 644.9 & \phantom{0}4.1& 1.40  & 1.35 & 72.1 & 3.0 & $3.15 \pm 0.2$ & $2.3 \pm 0.1$  & - \\
    cntH$\alpha$ - 28 Jan 15 & 50.534 & 11.5 & 1.2 x 36 & 644.9 & \phantom{0}4.1 & 1.25 & 0.93 & 59.4 & 9.0 & $5.45 \pm 0.9$ & $3.6 \pm 0.5$ & - \\
    cnt820 - 10 Dec 14 & \phantom{1}2.589 & \phantom{0}4.8 & 1.2 x 30 & 817.3 & 19.8 & 1.35 & 1.1\phantom{0} & 58.3 & - & - & - & ND1\\
    cnt820 - 10 Dec 14 & \phantom{1}2.584 & 14.4 & 1.2 x 30 & 817.3 & 19.8 & 1.35 & 1.2\phantom{0} & - & - & - & & sat.\\
\hline                                             
\end{tabular}
\tablefoot{The average Julian date and the total integration time of the observations are given in columns 2 and 3.
DIT is the exposure time of the individual frames and NDIT is the number of frames per exposure. Each cycle (DIT $\times$ NDIT) was repeated four times
to obtain $+Q$, $-Q$, $+U$, and $-U$ frames and the whole cycle was repeated several times for each filter and epoch until the total exposure time was reached.
AM and $\theta$ are the airmass and the visible seeing respectively.
$\lambda_\circ$ and $\Delta \lambda$ are the central wavelength and the full-width at half maximum of the filters used. FWHM$_\star$ is the azimuthally-averaged
full-width at half maximum observed for R~Dor. Peak PD, IPF, and MPF (defined in Section \ref{sec:obsConst})
stand, respectively, for the peak value of the polarisation degree, the integrated polarised fraction,
and the maximum polarised fraction (not given for cnt820,
see Section \ref{sec:pol-light}). In the last column, ND1 is the neutral density filter used and sat. indicates that the CCD is saturated.}
\end{table*}


\subsection{Data acquisition and data reduction}
\label{sec:data}

R\,Dor was observed with ZIMPOL during the SPHERE science verification time using three filters, V, cntH$\alpha$, and cnt820.
The observations were taken in two epochs, 10 and 11 of December 2014 (V, cntH$\alpha$, and cnt820)
and 28 of January 2015 (cntH$\alpha$).
The total integration times for each filter and epoch are given in Table \ref{tab:obs}. Observations using
filter cnt820 were done both with and without a neutral density (ND) filter. In the images obtained without
the ND filter, the star saturated the detector. When the first-epoch cntH$\alpha$ images were taken
the seeing was too high ($>1.2''$) for an optimal behaviour of the instrument,
therefore these data have to be interpreted with care.

The observations of R\,Dor resulted in individual data cubes, containing the frames recorded with the 
two cameras of the instrument (both equiped with the same filter). We processed these cubes individually
using the data reduction pipeline of the instrument, in its pre-release version 0.14.0\footnote{Downloadable
from {\em ftp://ftp.eso.org/pub/dfs/pipelines/sphere/}}. Each cube produced Stokes $+Q$, $-Q$, $+U$
and $-U$ frames for the two cameras, together with intensity frames $I_Q$ and $I_U$. We then aligned and de-rotated
the resulting average frames using custom {\tt python} routines. We adopt a pixel scale of
$3.602 \pm 0.011$\,mas\,pix$^{-1}$ and a position angle of the vertical axis with respect to North of
$357.95 \pm 0.55\,\deg$ (Ch. Ginski, in prep). Recently, new direction-dependent corrections to
the pixel scale of less than 0.5\% have been determined (see Ginski et al, in prep). Since these are much
smaller than the uncertainties from choosing the central pixel (see Section \ref{sec:obsConst}) for the region we consider, we have not included them.

We deconvolved the total intensity images (only) using the Lucy-Richardson (L-R) algorithm implemented in the IRAF\footnote{{\em http://iraf.noao.edu}}
software package.
The point-spread function (PSF) reference images of $\psi^2$\,Ori were
taken on the night of 31 March 2015, under comparable seeing conditions ($\sigma=0.9\arcsec$) as the
R\,Dor observations. For the cnt820 filter, we adopted the PSF observation in the TiO717
filter (with $\lambda_\circ = 716.8\,\mu$m and $\Delta \lambda = 19.7\,\mu$m) as no other PSF observation in the cnt820 filter was available. We stopped the
L-R deconvolution after 80 iterations, as the deconvolved images do not show a significant evolution in
additional processing steps. Since the PSF images were not acquired simultaneously to the images of R~Dor, we only show the deconvolved images
to illustrate what the underlying source morphology might be. All the quantities that we present and model were extracted from the observed images
and not from the deconvolved ones.

\subsection{Observational results}

\subsubsection{The stellar photosphere}

\begin{figure}[t]
   \centering
   \includegraphics[width= 9cm]{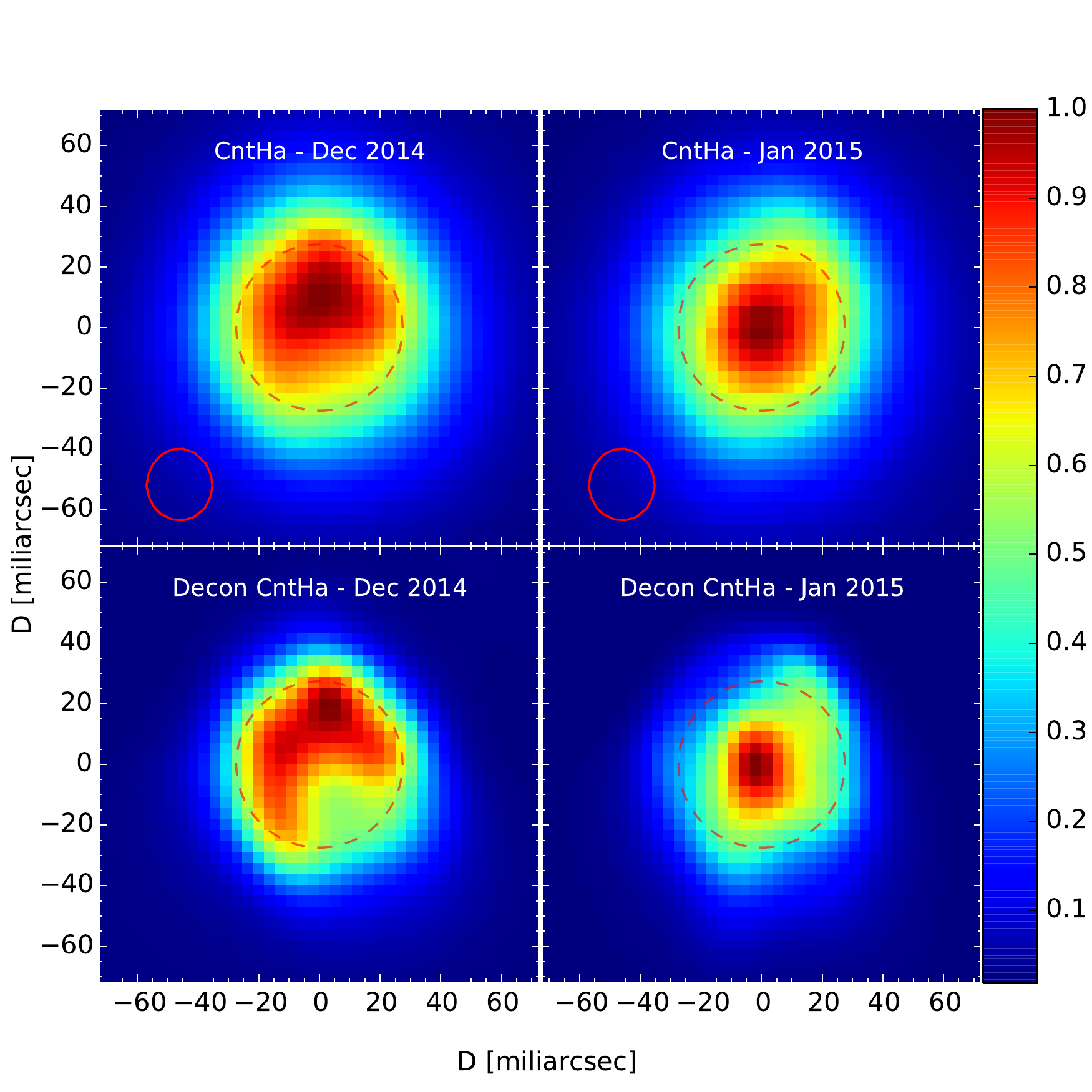}
      \caption{{\it Top panels}: total intensity normalised to unity as observed using ZIMPOL in cntH$\alpha$ on 10-Dec-2014 and 28-Jan-2015.
      {\it Bottom panels}:
      the corresponding deconvolved images, again normalised to unity. The dashed red circles show the size of the stellar disc
      derived by \cite{Norris2012} from observations in the near-infrared (also shown in Figs. \ref{fig:filters}, \ref{fig:filters_contours}, \ref{fig:polangle},
      and \ref{fig:polDeg}). The full red circles show the FWHM of the PSF used as reference.}
         \label{fig:cntHa}
   \end{figure}

We first discuss the total intensity (Stokes $I$) images.
The high spatial resolution achieved with SPHERE/ZIMPOL (${\approx 20\,{\rm mas}}$) allows us to resolve the stellar disc of R~Dor
in the total intensity images (Figs. \ref{fig:filters} and \ref{fig:cntHa}).
The images acquired in the first epoch reveal a very asymmetrical source, with a horseshoe-shaped morphology both in V and in cntH$\alpha$.
In these images, more emission arises from
the northern hemisphere and there is a region of low surface brightness in the south-west. We find an azimuthally-averaged
full-width at half maximum (FWHM) in V and cntH$\alpha$ in this epoch of $\approx 71\,$mas (see Table \ref{tab:obs}).
This asymmetric morphology is not as prominent in the cnt820 images and the source is also smaller at this wavelength, with FWHM $ \approx 58.3\,$mas.

The surface brightness distribution of R\,Dor in cntH$\alpha$ changes between the two epochs of observation (Fig. \ref{fig:cntHa}).
In Jan 2015, emission is concentrated in the central region and the intensity distribution is more axisymmetric,
different from what is seen in the first epoch. In the second epoch the disc of R~Dor has a FWHM $\approx 59.4\,$mas in cntH$\alpha$.
For reference, the 48 days that separate the two epochs correspond to slightly more than one-fourth of the
shortest pulsation period of R\,Dor (175 days).

The values of the FWHM
measured in the second epoch in filters cnt820 and cntH$\alpha$ are
comparable to the stellar radius for a uniform disc obtained by \cite{Norris2012} of $\approx 27.2$ mas, from observations
that probe the stellar continuum in the near-infrared. However, the values of the FWHM in the first epoch in V and in cntH$\alpha$
are significantly larger than that (see Table~\ref{tab:obs}).

A strong dependence of the size of the stellar disc on wavelength is a known feature of AGB stars. For instance, the measured uniform-disc diameters
in the near-infrared is found to correlate with molecular spectral bands of CO and H$_2$O \citep[see, e.g.,][]{Wittkowski2008,Woodruff2009}.
In the visible wavelength range, TiO is expected to be the main source of opacity for a late-type M star as R~Dor.
This molecule is found to dominate the spectrum in the wavelength
range of V and cntH$\alpha$, but its opacity is lower in the wavelength range of filter cnt820.
For comparison, molecular contribution is expected to be very
small in the wavelengths
that \citeauthor{Norris2012} performed their observations.
For examples of spectra of late-type stars with band identification, see, e.g., \cite{Lancon2000}.

\cite{Jacob1997} measured the diameter of R~Dor in the pseudo-continuum region around 0.82\,$\mu$m and in a TiO absorption band at
0.85\,$\mu$m and found the stellar diameter to be 20\% larger in the TiO band. Follow-up observations \citep{Jacob2004}
between 0.65\,$\mu$m and 0.99\,$\mu$m confirmed that the stellar disc size increases in the TiO bands.
\cite{Ireland2004} found asymmetries in the stellar disc of R~Dor in the same wavelength range observed by \cite{Jacob2004}.
The authors, however, could not determine whether these asymmetries were caused by variations in molecular excitation or in the light scattered
by dust grains. Since we find that the size and the morphology of the stellar disc change considerably while the polarisation degree does not decrease
significantly (see Section \ref{sec:pol-light}), we conclude that the difference in FWHM, the asymmetries, and the variation in morphology
are mainly caused by variability of molecular (TiO) opacity.

The variability in TiO opacity can be caused by changes in the gas density and/or in the molecular abundance or excitation.
Hydrodynamical models calculated
with the code CO5BOLD \citep{Freytag2013} for a star with a few solar masses show that convective motions can produce large-scale density variations
on time scales comparable to the one we find\footnote{See
\url{http://www.astro.uu.se/~bf/} for the model results.}. However,
these models lack a realistic wavelength-dependent radiative transfer that takes into account molecular opacity.
As molecular excitation, and hence the surface brightness of the stellar disc, is significantly affected by density variations, by episodic dissipation of energy
carried by shocks, and by variations in the stellar radiation field,
a quantitative comparison between observations and such models is not yet possible.

 \begin{figure}[t]
   \centering
   \includegraphics[width= 9cm]{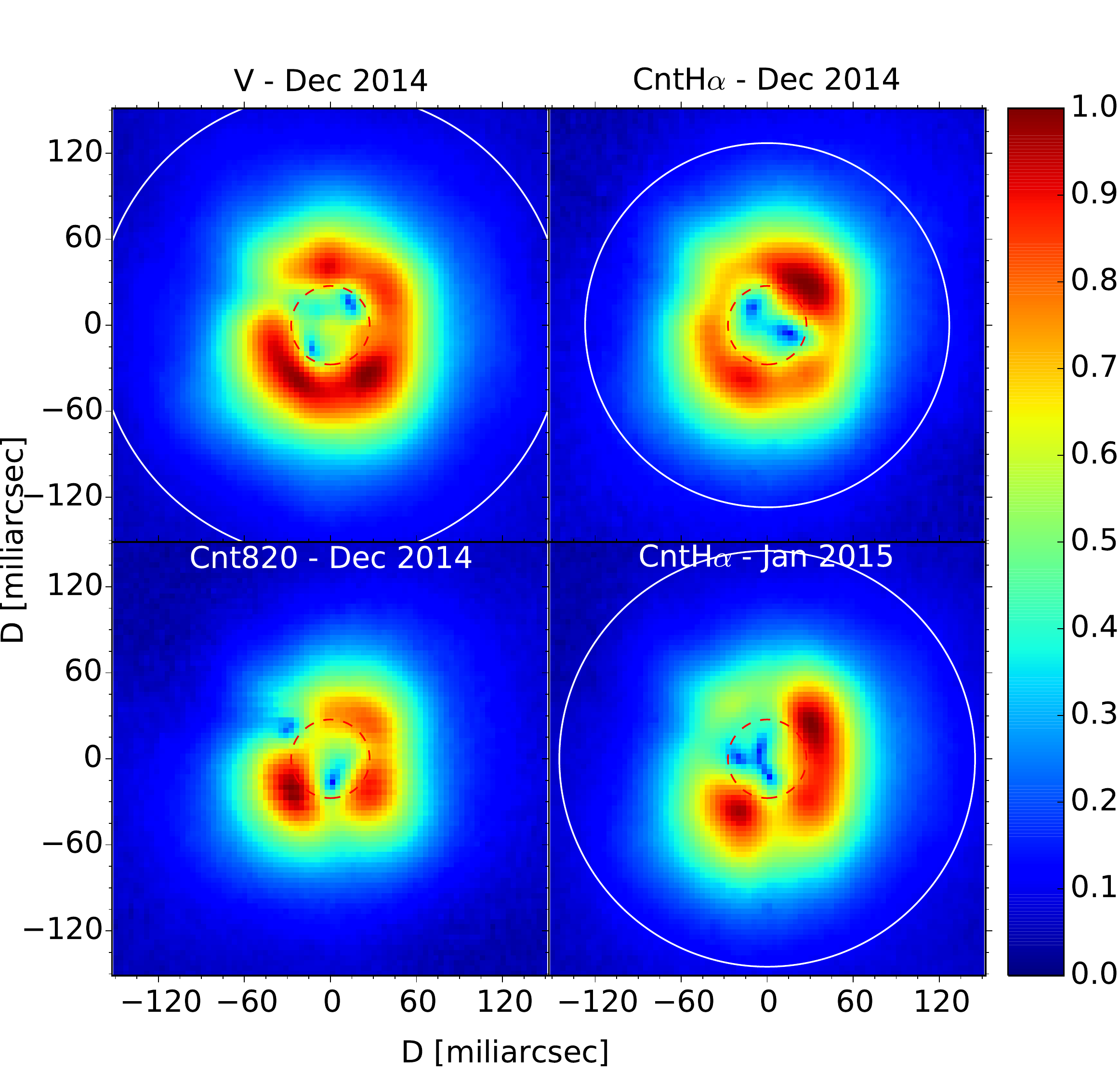}
      \caption{Polarised intensity seen in
      V, cntH$\alpha$, and in cnt820 on 11-Dec-2014 and in cntH$\alpha$ on 28-Jan-2015 normalised to the peak value,
      shown with a square-root scaling. The dashed red circles show the size of the stellar disc
      derived by \cite{Norris2012} from observations in the near-infrared (also shown in Figs. \ref{fig:filters}, \ref{fig:cntHa}, \ref{fig:polangle}, and \ref{fig:polDeg}).
      The white circles (not shown for cnt820, see Section \ref{sec:pol-light}) mark the region where we find the polarised light produced in the envelope
      to dominate over instrumental effects.}
         \label{fig:filters_contours}
   \end{figure} 

\subsubsection{The polarised light}
\label{sec:pol-light}

\begin{figure}[t]
   \centering
   \includegraphics[width= 9cm]{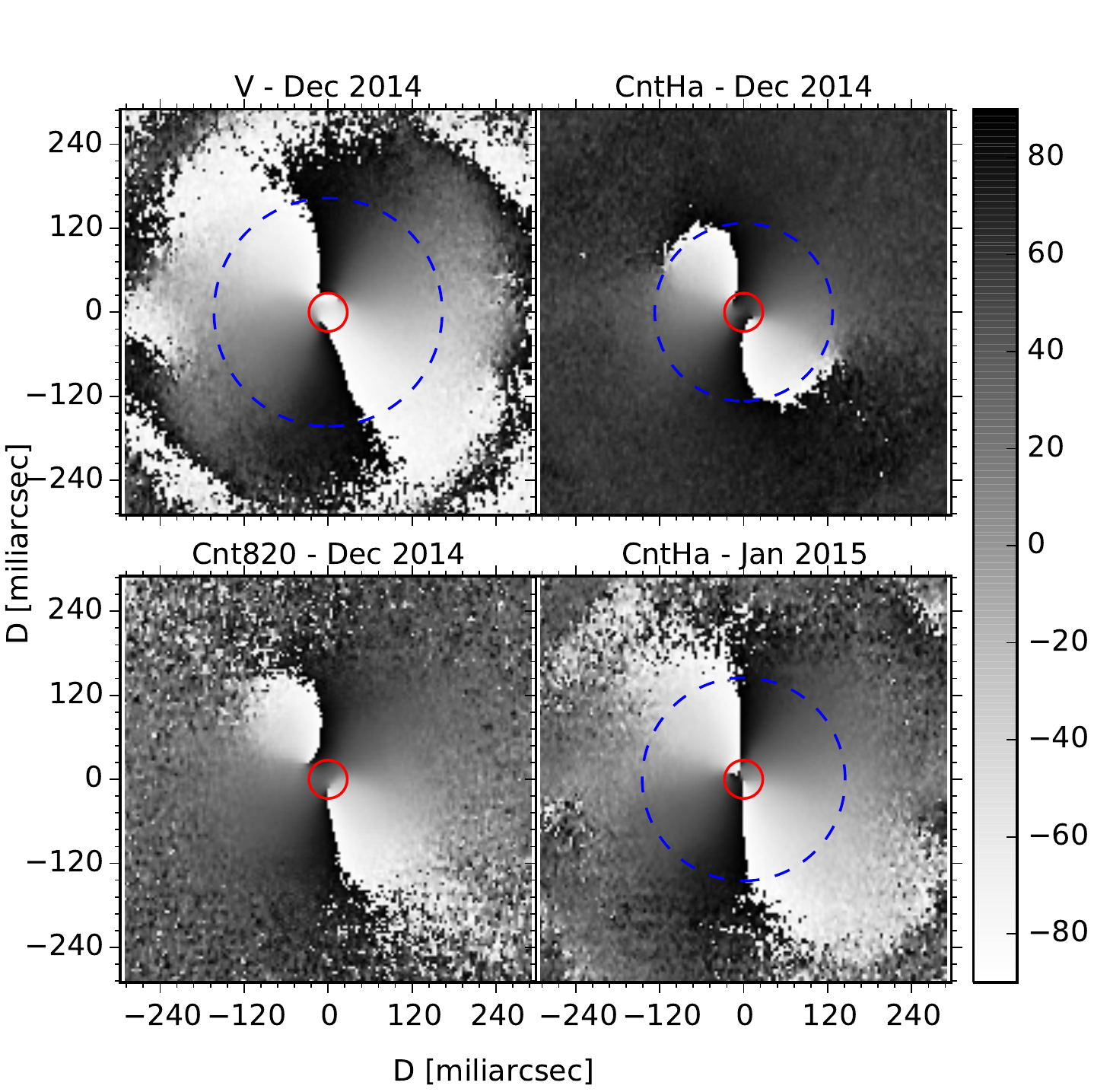}
      \caption{Direction of the measured polarisation vector for the four images given in degrees relatively to the north direction.
      The red circles show the size of the stellar disc derived by  \cite{Norris2012} from observations in the near-infrared (also shown in Figs. \ref{fig:filters}, \ref{fig:cntHa},
      \ref{fig:filters_contours}, and \ref{fig:polDeg}).
      The dashed blue circles enclose the regions where we find the polarised intensity produced in the envelope
      to be a factor of three larger than the expected instrumental polarisation (not shown for cnt820, see Section \ref{sec:pol-light}).}
         \label{fig:polangle}
   \end{figure}

Polarised light thought to be produced by scattering of stellar light off dust grains
was detected from around the central star (see Fig. \ref{fig:filters_contours}).
The observed polarisation vectors are in the plane of the sky and tangential to a circle centred on the star,
as expected for grains distributed in a circumstellar envelope illuminated by a central star.
Fig. \ref{fig:polangle} shows that the directions of the observed polarisation vectors 
match this expected behaviour up to distances of about 160 mas in the images obtained in the V filter,
about 130 mas in the first-epoch cntH$\alpha$ image, and about
145 mas in the second-epoch cntH$\alpha$ image.
At these radii the polarisation signal disappears in the noise.
We did not include the observations in cnt820 in this analysis because these
were either saturated or had to be acquired with a neutral density (ND) filter.
The region where the detector was saturated includes the inner rim of the ring from where polarised light is seen
and the ND filter can introduce uncalibrated instrumental polarisation.

The region from where we see polarised light is very similar for V and cntH$\alpha$ in the first epoch, but the brightness distribution is somewhat different.
In V the polarised intensity is slightly more concentrated in the south part of the image (60\%), while in cntH$\alpha$ both hemispheres show roughly the same
emission with 53\% of the polarised intensity originating from the southern hemisphere.
There is no obvious correlation between the directions of maximum or minimum polarised intensity and those with maximum or minimum total intensity.
Interestingly, although the total intensity distribution in cntH$\alpha$ changes drastically from one epoch to the other, the polarised intensity distribution does not
change significantly. The departure from axisymmetry is stronger in Jan 2015 but the location of the maxima and minima of the
polarised flux does not change much between the two epochs.

We divided the image in octants to facilitate our analysis and to minimise errors introduced by asymmetries of the dust envelope.
Octant 1 is limited by the north and north-east directions and the numeration follows clock-wise.
The value of the polarised intensity changes considerably for the
different octants. In Table \ref{tab:octs} we list the fraction of the polarised light measured per octant per filter. 
The ratio between the intensity from the octants with maximum and minimum polarised intensity is roughly 2.3 for V and cntH$\alpha$ in the first epoch and
3.7 for cntH$\alpha$ in the second epoch.

   \begin{table}
\caption{Fraction of the polarised flux arising from the different octants for the image in V and in cntH$\alpha$.}
\label{tab:octs}      
\centering                                      
\begin{tabular}{c | c c c}         
\hline                        
& \multicolumn{3}{  c  }{Fraction of polarised flux} \\
Octant & V & cntH$\alpha$ Dec & cntH$\alpha$ Jan \\
 & [\%] & [\%] & [\%]\\
\hline                                   
1 & 11\phantom{.5} & 15.5 & 13\phantom{.5} \\
2 & 11\phantom{.5} & 13.5 & 16\phantom{.5}\\
3 & 13\phantom{.5} & \phantom{1}9\phantom{.5} & 14.5 \\
4 & 17.5 & 13.5 & 14.5 \\
5 & 16\phantom{.5} & 17.5 & 18.5 \\
6 & 13.5 & 13.5 & 11.5 \\
7 & \phantom{1}7.5 & \phantom{1}7.5 & \phantom{1}5\phantom{.5}\\
8 & 10.5 & 10\phantom{.5} & \phantom{1}7\phantom{.5} \\
\hline                                             
\end{tabular}
\end{table}

 \begin{figure}[t]
   \centering
   \includegraphics[width= 9cm]{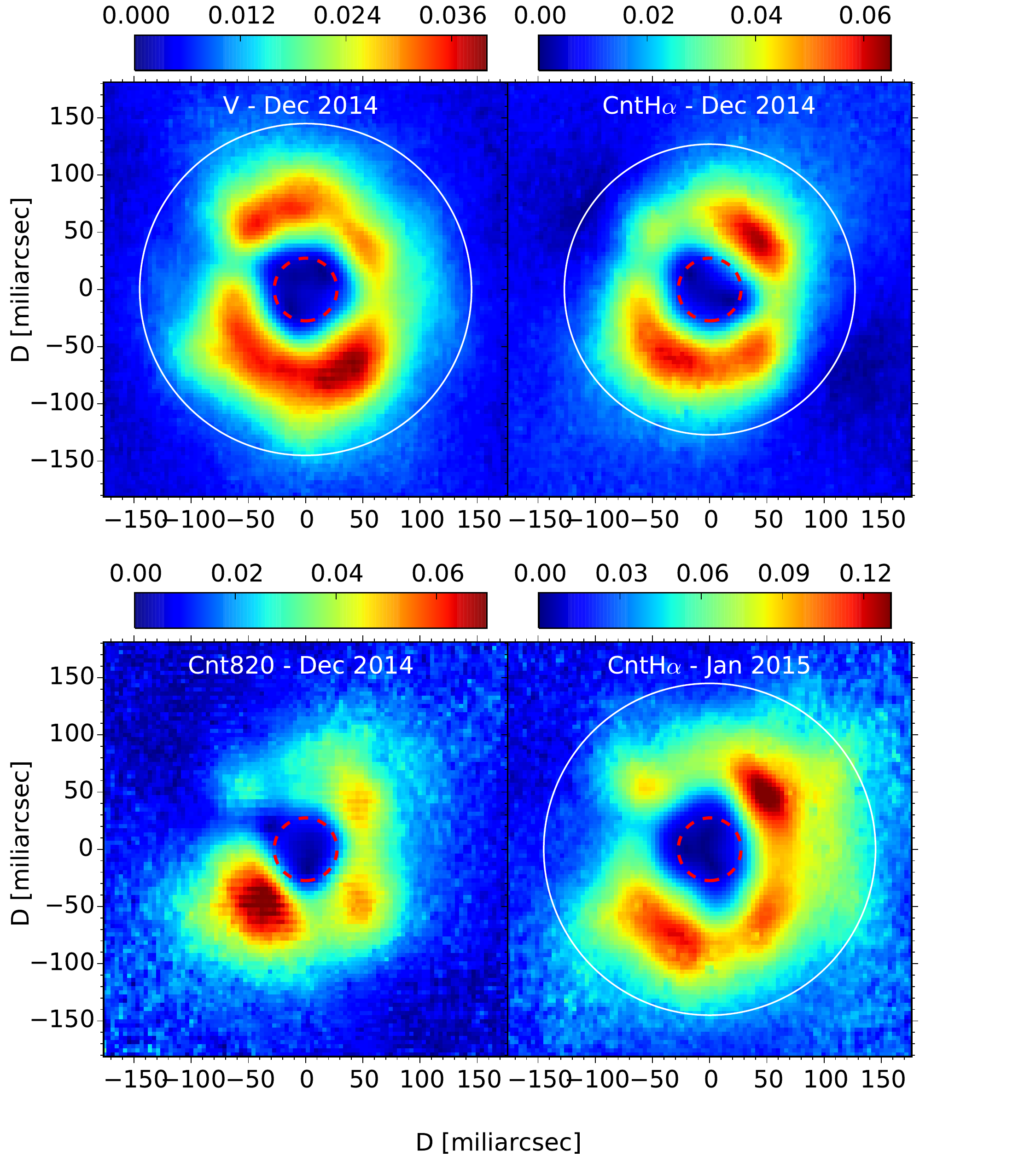}
      \caption{Polarisation degree for the four images.
      The red dashed circles show the size of the stellar disc derived by  \cite{Norris2012} from observations in the near-infrared (also shown in Figs. \ref{fig:filters}, \ref{fig:cntHa},
     \ref{fig:filters_contours}, and \ref{fig:polangle}).
      The dashed blue circles enclose the regions where we find the polarised intensity produced in the envelope
      to be a factor of three larger than the expected instrumental polarisation (not shown for cnt820, see Section \ref{sec:pol-light}).}
         \label{fig:polDeg}
   \end{figure}

\section{Analysis and modelling of the polarised light}

We now focus on our modelling efforts to derive the density radial profile of the grains that produce the scattered light.
 
\subsection{Modelling approach}
\label{sec:approach}
We calculated spherically symmetric models using the continuum radiative-transfer code MCMax \citep{Min2009}.
The code calculates the direction-dependent scattering of radiation by dust grains and outputs
images of the Stokes parameters.
We convolved the Q and U images produced by the models with the PSF images. This typically caused a decrease by a factor of 1.5
of the output integrated polarised flux. This is because in the Q and U images the negative and positive lobes from a symmetrical envelope
can overlap when the resolution is lowered. This leads to polarised signal in the images to cancel out. The poorer the angular resolution, the larger is the effect.
The convolved Q and U images were then combined to obtain the polarised intensity for each model.

We calculated the radial profile of the polarised intensity from the observations for each octant and compared these profiles independently to the models.
We considered two different envelope structures: with dust grains distributed in a thin halo with constant density (model with no outflow), and
with a dust density gradient given by $\rho(r) = \rho_\circ\, (R_\circ/r)^{n}$, $R_\circ$ being the inner radius of the dust envelope and $\rho_\circ = \rho(R_\circ)$.
We varied $R_\circ$ between 1.2~$R_\star$ and 2.0~$R_\star$ for both envelope structures and the halo thicknesses between
0.25~$R_\star$ and 1.5~$R_\star$ for the thin-halo scenario, where $R_\star = 27.2\,$mas \citep{Norris2012}.

We use amorphous Mg$_2$SiO$_4$ grains with the optical constants
obtained by \cite{Jager2003}, as this dust species is one of the main candidates for driving the winds of oxygen-rich AGB stars \citep[see, e.g.,][]{Bladh2012}.
We have also experimented with optical constants for Mg$_2$SiO$_3$ and Al$_2$O$_3$, as discussed in Section \ref{sec:dustSpecies}.
To calculate the absorption opacities and the direction-dependent scattering properties,
we approximate the actual size distribution of particle sizes by a distribution of hollow spheres \citep{Min2003}.
We consider the radii, $a$, to be given by the standard distribution for grains in the interstellar medium
$n(a) \propto a^{-3.5}$ \citep{Mathis1977}. The minimum, $a_{\rm min}$, and maximum, $a_{\rm max}$, grain radii of the distribution were varied to
fit the observations (see Section \ref{sec:mass}). Our fits are only sensitive to grains with $a \gtrapprox 0.1\,\mu$m,
since smaller grains do not provide significant scattering opacity. For distributions with $a_{\rm min} = 0.01\,\mu$m and $a_{\rm max}$ between
0.2\,$\mu$m and 0.5\,$\mu$m, the amount of mass in grains with $a \gtrapprox 0.1\,\mu$m is roughly between 40\% and 70\%.

\subsection{Observational contraints}
\label{sec:obsConst}

Our models only consider opacity due to dust grains, while molecular absorption may also be very important close to the star.
Neglecting this source of opacity will probably cause us to overestimate the inner radius of the dust envelope.
This is because photons scattered close to the star, where the gas densities are high, have a larger chance
of being absorbed before escaping the envelope. Hence, the inner radius of the dust envelope would appear larger.
 
Instrumental polarisation may also affect the observations, though, data reduction procedures
reduce it to about 0.5\%. This residual instrumental polarisation mainly affects the data at projected distances $> 100$\,mas
from the central source. To minimise the errors introduced, we only analysed the polarised light from
the inner region up to 127 mas from the star, where the polarisation degree is larger than
1.5\% and where we see the expected behaviour of the polarisation angles for the image in V and the two images in cntH$\alpha$.

We fit our models to the integrated polarised fraction (IPF), the maximum polarised fraction (MPF), and the radial profile
of the polarised intensity. We define the IPF as
the polarised intensity integrated within a radius of 127 mas divided by the total intensity integrated over the same region.
The MPF is the maximum value of the polarised fraction per octant integrated within a radius of 127 mas divided by the total intensity integrated
over a circle with a 127 mas radius.
These observed quantities are given in Table \ref{tab:obs}.
The errors given are the 1-$\sigma$ uncertainties derived from the uncertainties on the polarised intensity.
We also compared our models to the radial profile of the total intensity to set upper limits on the dust densities.

 \begin{figure*}[t]
   \centering
   \includegraphics[width= 18cm]{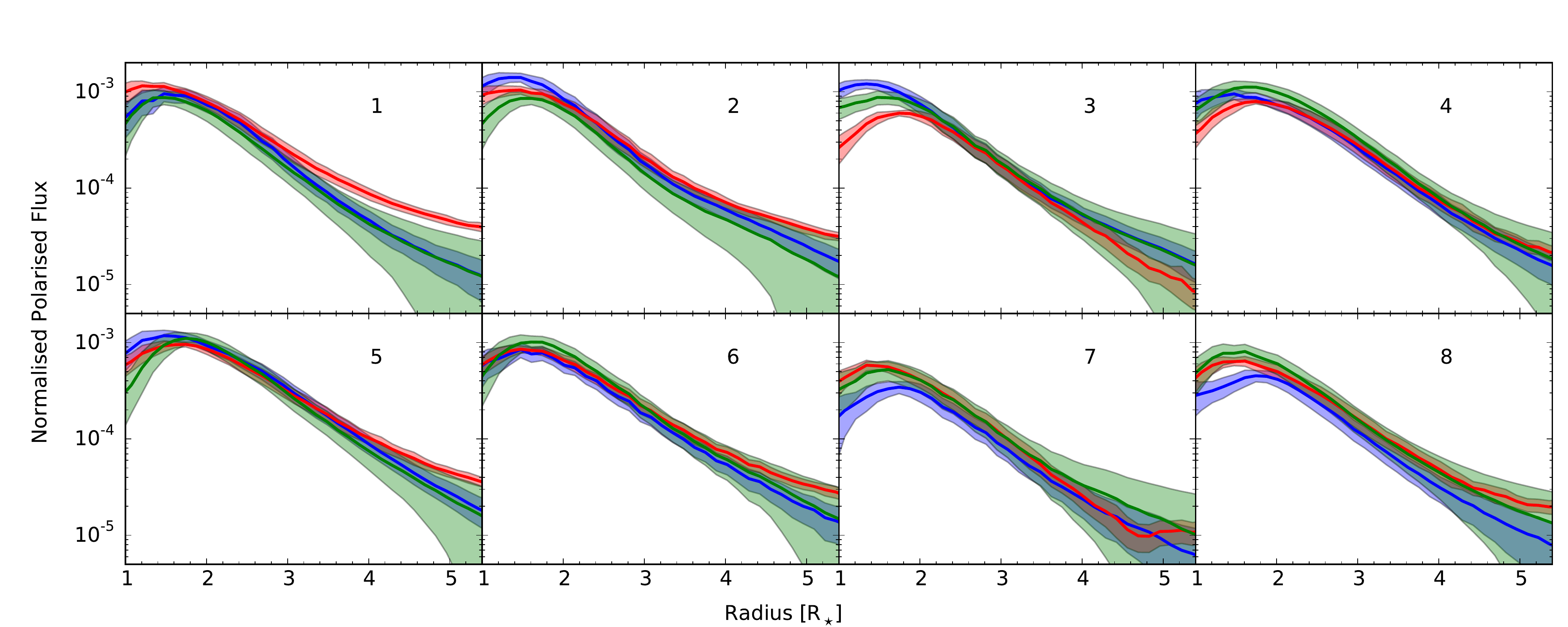}
      \caption{Comparison between the radial profile of the polarised intensity for the different octants obtained from the observations in the first epoch in V (green line) and
      in cntH$\alpha$ (red line), and in the second epoch in cntH$\alpha$ (blue line). The profiles were normalised using the value of the integrated polarised
      intensity for each filter. The octants are identified in each panel (see Section \ref{sec:pol-light}).
      The filled region shows the three-$\sigma$ errors from combining the uncertainty given by the ESO pipeline with that
      from choosing the central pixels.}
      \label{fig:radialP}
   \end{figure*}

The IPF in the image in V is significantly smaller than those in cntH$\alpha$ in both epochs.
The polarised signal in the first-epoch cntH$\alpha$ image can be affected by the not-ideal conditions
when the data were taken. Specifically, the Strehl ratio in that image is different from that in the reference
PSF ($\psi^2$~Ori), which was taken under good sky conditions. This can cause more instrumental polarisation to be produced,
affecting the IPF we measure. This is specially important in the regions where the polarisation degree is low ($\lesssim 1\%$).
The effect of the worse conditions can be seen in the images of the direction of the polarisation
vectors (Fig. \ref{fig:polangle}). The first-epoch image in cntH$\alpha$ shows the signal from circumstellar polarisation up to
a smaller distance than the second-epoch image. This would not be expected if both images had been taken under equal conditions
because in the first-epoch exposure time was longer.
Nonetheless,
decreasing the radius of integration from 127 mas to 110 mas only reduces the
value we obtain for the IPF by a factor of 1.01. Therefore, the effect of the worse sky conditions on the IPF are not significant.
Moreover, the observations in Jan 2015 also show a high IPF in cntH$\alpha$. This leads us to conclude that the higher IPF in cntH$\alpha$ when
compared to that in V is a real feature of the inner envelope of R~Dor. However, we were not able to reproduce this wavelength dependence of the IPF
with our models, which always show higher IPF in V or at most similar values of the IPF between the two filters.

An important consideration is that the measured IPF can be suppressed by molecular absorption. This is because the scattered photons travel longer through
the envelope than photons that do not interact and, hence, have a higher probability of being absorbed by molecules.
This causes the IPF produced by scattering off dust grains to decrease if the molecular absorption opacity increases in a given wavelength.
Given these considerations, it is more likely that polarised photons in the V band are absorbed by molecules than artificially created in cntH$\alpha$
because of instrumental polarisation.
Therefore, we consider the values measured in cntH$\alpha$ as more representative of the IPF produced by the dust.
This approach also guarantees that we do not underestimate the dust densities we derive.
Hence, we fit our models to the uncertainty-weighed-averaged IPF in cntH$\alpha$ from both epochs combined, of $2.35 \pm 0.1\%$.

In order to calculate radial profiles, we need to define a central pixel but the complex morphology of the source makes this determination difficult.
For reference, the FWHM of R~Dor in the total intensity images in V is of roughly 20 pixels (72 mas,
see Table \ref{tab:obs}).
We have chosen the central pixel in the images in V as the one more closely equidistant from
the peak of polarised light for different azimuths.
The central pixels in the other filters were chosen by establishing the best match
when overlaying the polarised-light images with those in V.
By following this approach, we find that the chosen central pixels differ from the centre of light by a distance of one (3.6 mas) or two pixels (7.2 mas)
for the images in V and in both epochs in cntH$\alpha$.
The pixels where the total intensity peaks differ from our chosen central pixel by a distance of between 3 (10.8 mas) and 4 pixels (14.4 mas) for the image in
V and the first-epoch image in cntH$\alpha$. For the second-epoch image in cntH$\alpha$, in which the stellar disc is more axisymmetric,
the intensity peaks on our chosen central pixel. These considerations support our approach for determining the central pixel.
The uncertainty caused by the choice of the central pixel on the derived radial profiles of the polarised intensity
was taken as the maximum difference between considering the chosen central pixels
or one of its eight neighbouring pixels. This approach overestimates the uncertainty on the slope of the radial profile of the polarised intensity.
The effect of considering different central pixels on the IPF and MPF is negligible compared to the uncertainty of the measurements.
 
 In Fig. \ref{fig:radialP} we compare the radial profiles for the eight octants in the images in V and in cntH$\alpha$. The profiles
 were normalised using the integrated polarised intensity for each filter.
 The radial profiles obtained from the three images are very similar and agree well within the uncertainties. The fact that the profiles from
 cntH$\alpha$ in the first epoch deviate slightly from the other two for $r \gtrsim 3\,R_\star$ can be attributed to the worse conditions when those
 observations were taken and to a higher instrumental polarisation.
 
\section{Results and discussion}

We were not able to determine the particle size from the wavelength dependence of the scattering.
This is because of the expected contribution of molecular opacity in the wavelength range of the observations
that can disrupt the wavelength dependence imprinted by scattering.
We infer that molecular opacity is important in the first epoch because of the higher IPF in V than in cntH$\alpha$, which
we are not able to reproduce with our dust model.

We fit the IPF of $2.35 \pm 0.1\%$
and the radial profiles within 127 mas (5\,$R_\star$, see Section \ref{sec:obsConst})
measured in the first epoch in V and in the second epoch in cntH$\alpha$.
This means that our results are limited to a small region around the star. For reference, material
moving outwards at the maximum gas expansion velocity, of 5.7\,km\,s$^{-1}$, would take roughly five years to cross the region we probe.
We also used the radial profile of the total intensity from the second-epoch observations in cntH$\alpha$ to set an upper limit on the
dust densities allowed by the models. Finally, we investigated whether a high value of $\rho_\circ$ is enough to reproduce the MPF.

\subsection{The integrated polarised fraction and the radial profile of the polarised intensity}
\label{sec:mass}

Optical-depth effects become important when
we consider grain size distributions with maximum radii, $a_{\rm max}$, between $0.2\,\mu{\rm m}$ and $0.5\,\mu{\rm m}$.
For these values of $a_{\rm max}$, our models require an average optical depth, $\tau_V$, $\gtrapprox 1$ to reproduce the observed IPF.
In the optically-thick regime our fit parameters, $\rho_\circ$, $R_\circ$, and $n$ cannot be determined independently.

Typically, the IPF we obtain from our models is larger the larger the dust mass within the region of integration (127 mas).
However, when $\tau_V \gtrapprox 2$, increasing $\rho_\circ$, and hence the dust mass, does not lead to a larger IPF
because of the effect of multiple scattering. Also, for $\tau_V \gtrapprox 1.5$, the peak of the radial profile of the polarised intensity
becomes broader and shifts to larger radii, in comparison to optically-thin models. This causes models with $\tau_V \gtrapprox 1.5$ to
require smaller values of $R_\circ$ and larger values of $n$, when compared to optically-thin models.
Our models with $\tau_V \lessapprox 1.5$ require $R_\circ = 1.45 \pm 0.1\,R_\star$ and $3.5 \lessapprox n \lessapprox 5$ (with an average best-fitting value of $n\approx 4.5$)
to fit the radial profile of the polarised intensity.
A model with $\tau_V \approx 3.5$ requires $R_\circ \approx 1.2$ and $n \approx 5.5$ instead.

\subsubsection{Constraints from the total intensity images}

\begin{figure}[t]
   \centering
   \includegraphics[width= 9cm]{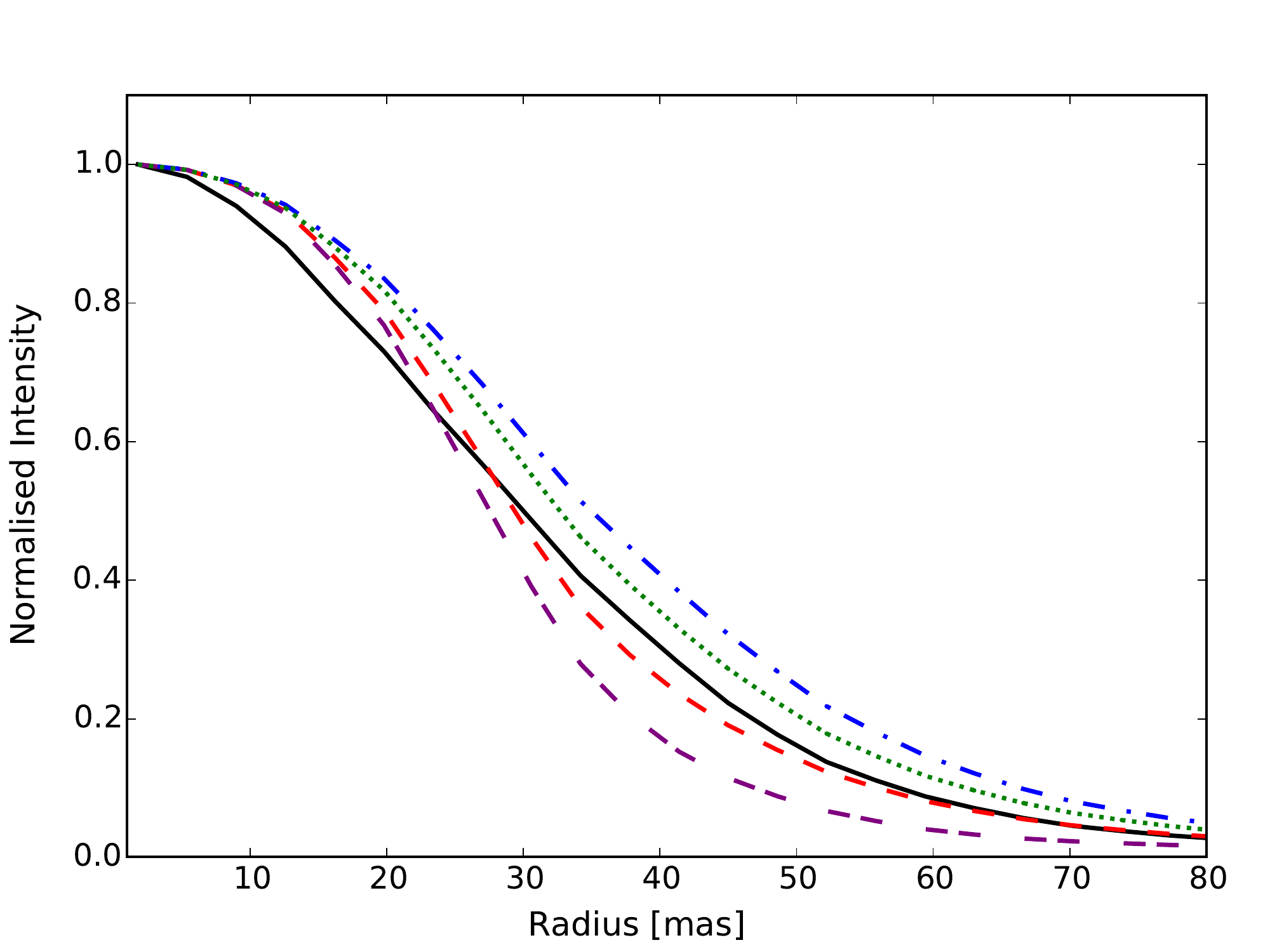}
      \caption{Observed azimuthally-averaged radial profile of the total intensity in the second-epoch cntH$\alpha$ images (full black line) compared to models
      with $\tau_V = 0.65$, $R_\circ = 1.5\,R_\star$, and $n=4$ (dashed red line),  $\tau_V = 1.3$, $R_\circ = 1.45\,R_\star$, and $n=4.5$ (dotted green line), and
      $\tau_V = 1.9$, $R_\circ = 1.4\,R_\star$, and $n=5$ (dotted-dashed blue line). The values of $R_\circ$ and $n$ for each of the dust models shown
      have been determined by fitting the radial profile of the polarised intensity.
For reference, we also show the radial profile of
      our model star with no dust envelope (double-dashed purple line).}
         \label{fig:totIntProf}
   \end{figure}
   
Although we do not attempt to fit the unpolarised surface brightness of R~Dor, we use the radial profile of the total intensity
to set an upper limit on the scattering optical depth of the envelope. The first-epoch images do not offer strong constraints because the stellar disc is seen to be very large in that epoch
both in V and in cntH$\alpha$. However, as shown in Fig. \ref{fig:totIntProf}, the second-epoch images in cntH$\alpha$ do offer important constraints.
Optically-thick models overpredict the total intensity for all distances from the star. Optically-thin models also overpredict the total intensity for $r\lessapprox 30\,$mas
but that is because our limb-darkened stellar model is not accurate enough to reproduce the brightness distribution of the stellar disc.
Since the polarised fraction in cntH$\alpha$ does not seem to decrease from the first to the second epoch, we conclude that this limit on $\tau_V$ applies to
both epochs. This reinforces the conclusion that
the variability and asymmetries in the total intensity images in the first epoch are caused by molecular opacity.
We conclude that models with scattering optical depths $\gtrapprox 1.2$ do not provide acceptable fits to the data.

Moreover, the value derived by \cite{Norris2012} for the inner radius of the dust envelope of R~Dor, of $R_\circ = 1.6\, R_\star$,
is in better agreement with what we obtain for models with $\tau_V \lessapprox 1.5$.
Since the scattering opacities in the wavelengths of the observations of \citeauthor{Norris2012} are roughly one order of magnitude lower
than those in the wavelength range of our observations, it is unlikely that their results are affected by high scattering optical depths in the envelope.
This reinforces the conclusion that optically-thin models are preferred.

\subsubsection{Larger grain sizes and different dust species}
\label{sec:dustSpecies}

We have tested the effect of increasing $a_{\rm max}$
to values between 0.6 $\mu$m and 1.0 $\mu$m. We find that such models produce lower values of the IPF for a given optical depth,
when compared to models with smaller values of $a_{\rm max}$. Therefore, the effects of high $\tau_V$
are stronger the larger $a_{\rm max}$, for $a_{\rm max} \gtrapprox 0.1\,\mu$m.

We have also experimented using optical constants for amorphous Al$_2$O$_3$ \citep{Koike1995} and MgSiO$_3$ \citep{Dorschner1995},
with all other parameters kept constant. We find IPF values that differ only by a few percent
from those obtained using Mg$_2$SiO$_4$. We conclude that the effect of the assumed grain species on our results is small.

\subsubsection{The radial profile of the dust density}
\label{sec:radialProf}

\renewcommand{\arraystretch}{1.2}

\begin{table}
\caption{The calculated reduced-$\chi^2$ values of the models shown in Fig.~\ref{fig:V-models} fit
to the radial profiles measured in the V filter for each octant.
These were computed using the normalised profiles of models and observations between 18~mas and 127~mas
to reflect the goodness of the fit to the shape of the observed radial profiles.
The lowest reduced-$\chi^2$ value for each octant is highlighted with boldface.}
\label{tab:chi2}      
\centering                                      
\begin{tabular}{c | c c c c c c c c} 
Model & \multicolumn{8}{c}{Octants} \\
 & 1 & 2 & 3 & 4 & 5 & 6 & 7 & 8 \\
\hline
$r^{-2}$ & 223 & 194 & 135 & 184 & 176 & 226 & 57 & 172 \\
$r^{-3}$ & 52 & 45 & 22 & 35 & 35 & 48 & 7.1 & 33 \\
$r^{-4}$ & 5.8 & 6.4 & {\bf 0.63} & {\bf 2.2} & 2.5 & 5.3 & {\bf 0.27} & 1.6 \\
$r^{-5}$ & {\bf 0.56} & {\bf 1.1} & 2.4 & 3.1 & {\bf 2.3} & {\bf 0.86} & 2.9 & {\bf 0.63} \\ 
Shell & 7.3 & 6.5 & 15 & 19 & 16 & 15 & 12 & 11 \\
\end{tabular}
\end{table}

\begin{figure*}[t]
   \centering
   \includegraphics[width= 18cm]{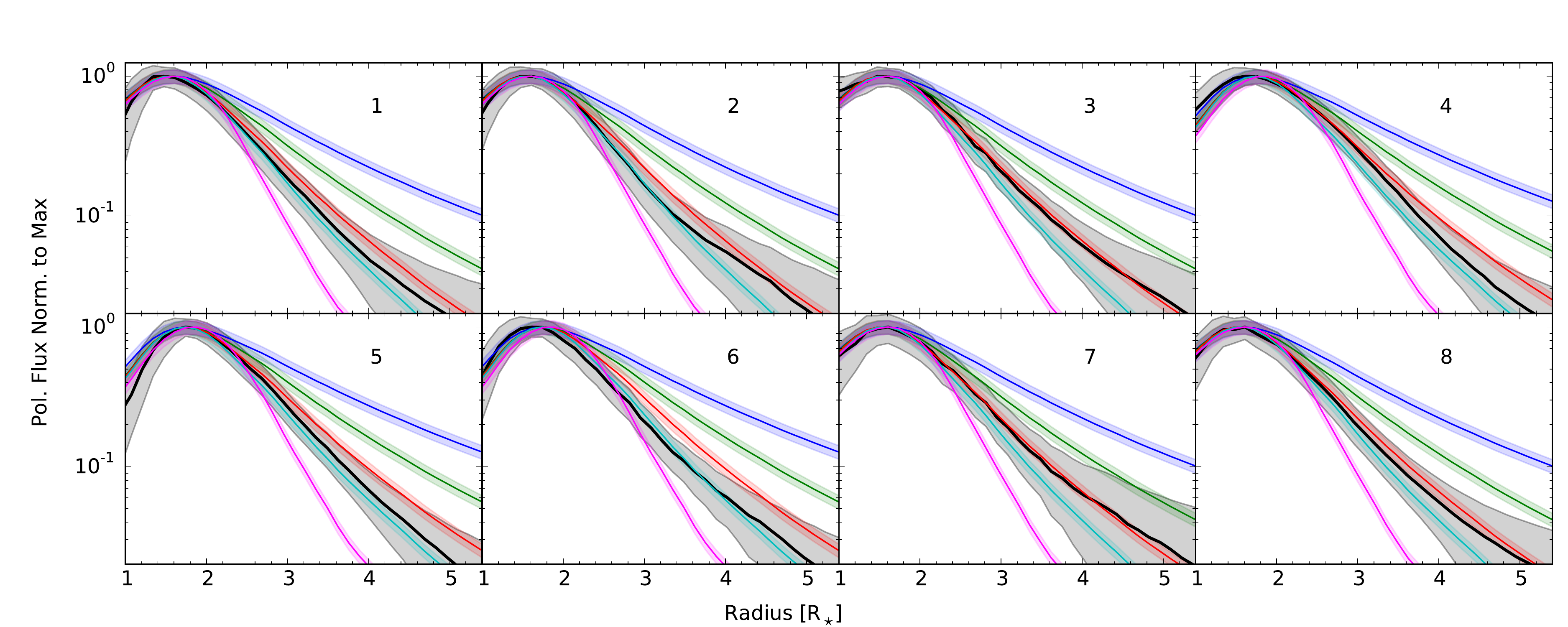}
      \caption{Comparison between the radial profile of the polarised intensity obtained from models and from the observations in V (black line)
      for the different octants. The fraction of the IPF arising from each octant is given in percents.
      Octant one is limited by the north and north-east directions and the numeration follows clock-wise.
      The grey-filled region shows the three-$\sigma$ errors from combining the uncertainty given by the ESO pipeline with that
      from choosing the central pixel. The different octants are identified by number in each panel (see Section \ref{sec:pol-light}).
      We show the best fit models for: a thin halo (pink line), $n=5$ (cyan line), $n=4$ (red line), $n=3$ (green line), and
      $n=2$ (blue line). The filled regions around the model lines show variation with direction introduced by
      the convolution with the PSF.}
         \label{fig:V-models}
   \end{figure*}

The best-fits to the observed radial profile measured in V are shown in Fig. \ref{fig:V-models}.
In Table~\ref{tab:chi2}, we show reduced-$\chi^2$ values for these same models fit to the normalised profile of the
observed polarised flux.
As can be seen, models with dust grains confined to a thin halo are not able to reproduce the observations
for $r \gtrapprox 2.5\,R_\star$,
while models with $n \lessapprox 3$ have dust density profiles that are too shallow and that overpredict the observed polarised intensity
also for $r \gtrapprox 2.5\,R_\star$.
The dust density profiles we find are not sensitive to the size of the particles we consider, as long as the optical depth in the visual,
$\tau_V$, does not vary.

The best-fitting values of $\tau_V$ and $\rho_\circ$ for models with different minimum and maximum grain sizes
are given in Table \ref{tab:res}.
The dust density profiles ($n\approx 4.5$) we obtain are steeper than that of a wind expanding with constant speed ($n = 2$).
This can be the result of
acceleration, of destruction of the dust grains in the observed region, of a decreasing mass-loss rate on short time scales,
or be caused by the density structure imprinted by consecutive
shock waves in the inner envelope, as predicted by 1-D wind-driving models
\citep[see, e.g.,][for gas density profiles from models for carbon-stars]{Hofner2003}.

\cite{Ireland2005} observed polarised light from two low-mass-loss rate and oxygen-rich AGB stars, R~Car and RR~Sco, and found that the data
was better reproduced by a model with a central star surrounded by a thin shell, instead of an outflow. If these sources also have steep density profiles in the inner wind,
the structure they interpreted as a thin shell might actually be the edge of a dust envelope similar to that of R~Dor.

Our models produce values of the scattered-light fractions in the near-infrared roughly a factor of 3.5 larger
than those reported by \cite{Norris2012} for R~Dor. It is not clear whether this is caused by variability of R~Dor
between the times of the two observations, by systematic errors in the data acquisition methods,
or by differences in the models used by us and by \citeauthor{Norris2012} to derive these quantities.

\subsection{The maximum polarised fraction}

We estimated the variation in $\rho_\circ$ required to reproduce the difference in the observed polarisation degree between the octants.
We are not able to reproduce the MPF in cntH$\alpha$ ($> 3$ \%) with models that have $ a_{\rm max}\gtrapprox 0.2\,\mu{\rm m}$ because the
model envelopes become too optically thick and do not reach such high polarised fractions.
By considering $a_{\rm max} = 0.1\,\mu{\rm m}$ we are able to fit the observed IPF with much lower maximum optical depths of $\approx 0.2$ and
we are, then, able to reproduce the MPF.

\cite{Norris2012} report grains of 0.3\,$\mu$m in the outflow of R~Dor. This value is in the range of grain radii for which we cannot reproduce the MPF.

\subsubsection{Tangential optical depths}
\label{sec:tang}

Given the considerable optical depths in the wavelengths we observe, asymmetries can cause tangential
optical depths in the envelope to be significantly different from those in our spherically symmetrical model. The tangential optical depths are measured
along the line-of-sight and in the direction of the observer between the plane of scattering and the outer edge of the envelope.
For lines-of-sight with lower tangential optical depths, more polarised photons
can escape. In this case, the MPF could be reproduced by an envelope with grains with $a_{max} \gtrsim 0.2\,\mu$m and
tangential optical depths varying with azimuth. In order for the polarised intensity from
one octant to be a factor of 1.5 higher than that of a spherically symmetric model,
the tangential optical depth would have to be 0.4 smaller. Given that the optical depths we find are of order unity, this would imply a decrease of
the tangential optical depth of roughly 40\% for the octants with maximum polarisation.
Since the polarised fraction would depend on both the tangential optical depth and on $\rho_\circ$
for each octant for non-spherically-symmetric envelopes
we have not explored this possibility further.

\begin{table}
\caption{Best-fit models for the radial distribution of scattering grains using different minimum and maximum grain sizes. $\rho_\circ$
is the density needed at $R_\circ$ to produce the IPF for a model
with $n=4.5$. $\tau_V$ is the radial V-band optical depth of the corresponding model. We also show the
range in $\rho_\circ$ and $\tau_V$ required to reproduce the observed variation of polarisation degree between the octants
with maximum and minimum values. The question marks indicate that we were not able to reproduce the MPF because the
models become too optically thick.}
\label{tab:res}      
\centering                                      
\begin{tabular}{c c c}         
\hline\hline                        
$a$ & $\rho_\circ$ & $\tau_{\rm V}$ \\
($\mu$m) & (g\,cm$^{-3}$) & \\
\hline                                   
0.01 - 0.05 & $(3.3 \pm 1.1) \times10^{-17}$ & $0.19 \pm 0.07 $ \\
0.01 - 0.1 & $(6.5 \pm 2.5) \times10^{-18}$ & $0.22 \pm 0.08 $ \\
0.01 - 0.2 & $(1.0^{+?}_{-0.45} \pm ) \times10^{-17}$ & $ 1.2^{+?}_{-0.5} $ \\
0.01 - 0.3 & $(7.2^{+?}_{-3.4} ) \times10^{-18}$ & $1.25^{+?}_{-0.6} $ \\
0.01 - 0.4 & $(6.0^{+?}_{-3.0} ) \times10^{-18}$ & $ 1.10^{+?}_{-0.45} $ \\ 
0.01 - 0.5 & $(6.3_{-3.2}^{+?}) \times10^{-18}$ & $ 1.15^{+?}_{-0.5} $ \\ 
0.3 - 0.5 & $( 3.0_{-1.25}^{+?} ) \times10^{-18}$ & $ 1.1^{+?}_{-0.45} $ \\ 
\hline                                             
\end{tabular}
\end{table}

\subsection{Comparison with ZIMPOL results for W~Hya}

\cite{Ohnaka2016} reported SPHERE/ZIMPOL observations of the O-rich AGB star
W~Hya that resolved the direct stellar emission and the region in the close circumstellar environment where polarised light is produced.
The authors find large grain sizes ($\sim~0.5~\mu$m) and small optical depths, of 0.1, in visual wavelengths.
These are very different from our preferred models for R~Dor, with optical depths in visible wavelengths of $\sim 1$.
Since the polarisation degree reported by \citeauthor{Ohnaka2016} for W~Hya is comparable to what we find for R~Dor and
the mass-loss rate of W~Hya \citep[of $\sim 1.3\times10^{-7}~{\rm M}_\odot$~yr$^{-1}$,][]{Khouri2014} is similar to that of R~Dor, the differences
between the results reported by \citeauthor{Ohnaka2016} and ours are not expected.

\citeauthor{Ohnaka2016} used a non-spherically symmetric radiative transfer model to reproduce the observations. This might
affect the derived optical depths because the tangential optical depths can be considerably different from a spherically symmetric model (see Section \ref{sec:tang}).
Moreover, the authors constrained the dust mass also taking into account the scattered light fractions reported by \cite{Norris2012} for W~Hya.
Therefore, the differences in the derived
optical depths might be related to the fact that we overpredict the scattered light fractions
given by \citeauthor{Norris2012} for R~Dor (see Section \ref{sec:radialProf}).
However, we are not able to determine the cause of these discrepancies and further investigation is needed.

We note that if our models underestimate the polarised flux for a given value of the optical depth of the envelope,
the dust densities we find will be overestimated.
Hence, the actual dust densities and the dust-to-gas ratio in the envelope will be smaller than the values we report.

\subsection{Dust grains as wind drivers}

We now investigate whether the grains we see around R~Dor are sufficient for driving the wind.
Wind-driving models indicate that the minimum dust-to-gas ratio ($d/g$) for a wind to develop is $d/g \approx 6 \times 10^{-4}$ \citep{Hofner2008} or
even $d/g \approx 3.3\times10^{-4}$ \citep{Bladh2015} \footnote{Both studies report silicon condensation fractions. We calculated the $d/g$ by adopting
solar silicon abundances \citep{Asplund2009}}. These models consider grains of a single radius, which is typically $\gtrapprox 0.2\,\mu$m.

We estimate an empirical upper limit for the $d/g$ at a radius $r$ in the envelope of R~Dor
by considering a lower limit for the gas density, ${\rho_{\rm gas}(r) \geqq \frac{\dot{M}_{\rm gas}}{4\pi r^{2}
\upsilon_\infty}}$, and the dust density profiles we find. We use the parameters commonly obtained
for the gaseous outflow of R~Dor of ${\dot{M}_{\rm gas} = 9\times10^{-8}\,{\rm M_\odot\,yr^{-1}}}$ and ${\upsilon_\infty = 5.7\,{\rm km\,s^{-1}}}$ \citep[see, e.g.,][]{KhouriPhD}.
This should give us a robust lower limit on the gas densities, since the gas expansion velocities are expected to be smaller than $\upsilon_\infty$
close to the star.

For the dust density profile, we use $n=4.5$ and the average density from the model with only large grains, ${\rho_\circ = 3\times10^{-18}\,{\rm g\,cm^{-3}}}$,
as this is the model more directly comparable to the single-grain-size models of \citeauthor{Hofner2008} and \citeauthor{Bladh2015}. This choice should not have
a strong effect on our results because, at a given distance from the star, all our best-fitting models have densities of grains with
$a\geqq 0.15\,\mu$m which are always smaller or very similar to
that of the model with only large grains.

We find $d/g \leqq 5\times10^{-3}$ at $r = 1.5\,R_\star$ and $d/g \leqq 2\times10^{-4}$ at $r = 5.0\,R_\star$.
Although the upper limit for the $d/g$ we derive at $r = 1.5\,R_\star$ does not provide strong constraints, the low value for the upper limit at $r = 5.0\,R_\star$
shows that at that radius the $d/g$ in the outflow of R~Dor is close to the limit of what wind-driving models require.
We note that our results can underestimate the dust densities if the polarisation efficiency of the grains is overestimated by our model.
However, this would imply higher scattering optical depths for the envelope, which our results disfavour.
Given the low expansion velocity and mass-loss rate of R~Dor, a low value of the $d/g$ is not unexpected.
However, if the steep radial profile of the dust density extends beyond 5\,$R_\star$, the $d/g$ would soon become
too low for the dust to provide the required opacity to drive the outflow.
This would be a problem because the maximum expansion velocity of the outflow of R~Dor
only equals the escape velocity of a 1\,M$_\odot$ star at $r \approx 35\,R_\star$, and hence
the radiation pressure force would have to be larger than the gravitational pull up to that distance.

\section{Summary and conclusions}

We have observed the oxygen-rich AGB star R~Dor using SPHERE/ZIMPOL on the VLT in three filters: V, cntH$\alpha$, and cnt820. Observations in cntH$\alpha$
were acquired in two epochs 48 days apart. The stellar disc is resolved in all observations and we are able to study asymmetries and variability in the star.
We find the total intensity distribution of R~Dor
to have a horseshoe-shaped morphology in the first epoch both in V and in cntH$\alpha$. In the pseudo-continuum filter cnt820 the stellar disc is smaller and
any departures from axisymmety are much less pronounced. In the second-epoch, taken 48 days later, the image in cntH$\alpha$ shows a source with a very different morphology.
Moreover, the stellar disc is significantly smaller than in the first-epoch image in the same filter and is comparable in size to what we see in cnt820, also in the first epoch.
We interpret these differences in size and morphology as being caused by variability in the excitation and/or density of TiO molecules in the extended atmosphere of the star.

We detect polarised light coming from a ring that encloses the central source and that spans a similar region in the three images we consider,
in V and the two epochs in cntH$\alpha$. However, the polarised intensity varies significantly with azimuth for the three images and the ratio of the polarised intensity between
two images also varies somewhat with azimuth. We find the integrated polarised fraction in V to be smaller than those in cntH$\alpha$ in the two epochs and the integrated
polarised fraction to increase in cntH$\alpha$ from the first to the second epoch. Our fits to the integrated polarised fraction and the radial profile of the polarised intensity
show that we see outflowing dust grains. Considering models with dust density profiles of the type $\rho(r) = \rho_\circ r^{-n}$, we find that the dust density decreases
much more steeply with radius than expected for a wind expanding at constant speed. This can be caused by the acceleration of the wind, but could also be explained by
destruction of the dust grains or by a varying mass-loss rate on short time scales.

We use our best-fitting dust models and literature values for the gaseous outflow to calculate upper limits for the dust-to-gas ratio as a function of radius. We compare the
limits we obtain to results from wind-driving models and we find that the upper limit we derive for the dust-to-gas ratio at $5\,R_\star$ is somewhat lower than
the minimum values required by such models for a wind to develop. Given the approximations we use for the grain model and the envelope structure, the upper limit we find
for the dust-to-gas ratio is consistent with the value found in wind-driving models. However, if the steep dust density gradient we derive extends to larger radii, the dust-to-gas
ratio would surely become
too small for the outflow to be driven. Given the low expansion velocity of the outflow of R~Dor, the wind only reaches the escape velocity of a one solar mass star about
$35\,R_\star$ from the central star. Therefore, if the grains we see are the main source of opacity that drives the wind,
we would expect a flattening in the power law of the dust density profile not much farther out than the maximum radius we probe.

Further investigation of the outflow of R~Dor will help to better understand how its wind is accelerated.
Particularly, deeper observations using ZIMPOL
can probe the dust density farther out in the envelope to better constrain the role of the scattering grains in driving the wind.
Similar studies for other AGB stars will show whether what we see for R~Dor is representative in any way and how the distribution of the dust in the inner wind changes
for different stars. This will help advance our knowledge of the driving of AGB outflows and of the AGB evolution in general.

\begin{acknowledgements}
This work was supported by the Swedish Research Council.
M.M. has received funding from the People Programme (Marie Curie Actions) of the EU’s FP7 (FP7/2007-2013) under REA grant agreement No. 623898.11.
W.V. acknowledges support from ERC consolidator grant 614264.
\end{acknowledgements}

\bibliographystyle{aa}
\bibliography{../bibliography_2}

\end{document}